\documentclass[12pt]{aastex631}
\usepackage{placeins,verbatim}
\usepackage{xcolor}
\usepackage{amsmath}

\usepackage{amsmath}
\usepackage{rotating}
\makeatletter
\usepackage[flushleft]{threeparttable}
\usepackage{threeparttable}
\usepackage{hyperref}
\usepackage{multirow}

\newcommand{\degree}{{}$^\circ$}

\newcommand{\HI}{\ion{H}{1}}
\newcommand{\ujy}{$\mu$Jy}
\newcommand{\pkg}[1]{\texttt{#1}}

\shorttitle{CHILES Con Pol-II: Radio Continuum Source Catalog}
\shortauthors{Gim et al.}

\graphicspath{{./}{figures/}}

\begin{document}

\title{The CHILES Continuum \& Polarization Survey-II: Radio Continuum Source Catalog and Radio Properties}

\correspondingauthor{Hansung B. Gim}
\email{hansung.b.gim@gmail.com}

\author[0000-0003-1436-7658]{Hansung B. Gim}
\affiliation{Department of Physics, Montana State University, Bozeman, MT 59717, USA}
\affiliation{Department of Astronomy, University of Massachusetts, Amherst, MA 01003, USA}
\affiliation{Eureka Scientific, 2452 Delmer Street, Suite 100, Oakland, CA 94602, USA}

\author[0000-0001-7095-7543]{Min S. Yun}
\affiliation{Department of Astronomy, University of Massachusetts, Amherst, MA 01003, USA}

\author{Nicholas M. Luber}
\affiliation{Department of Astronomy, Columbia University, Mail Code 5247, 538 West 120th Street, New York, NY 10027}

\author[0000-0003-3168-5922]{Emmanuel Momjian}
\affiliation{National Radio Astronomy Observatory, P.O. Box O, Socorro, NM 87801, USA}

\author[0000-0001-7996-7860]{D. J. Pisano}
\affiliation{Department of Astronomy, University of Cape Town, Private Bag X3,
Rondebosch 7701, South Africa}

\author[0000-0001-9662-9089]{Kelley M.~Hess}
\affiliation{Department of Space, Earth and Environment, Chalmers University of Technology, Onsala Space Observatory, 43992 Onsala, Sweden}

\author[0000-0002-9627-7519]{Julia Blue Bird}\footnote{Julia Blue Bird was a Jansky Fellow of the National Radio Astronomy Observatory.}
\affiliation{National Radio Astronomy Observatory, P.O. Box O, Socorro, NM 87801, USA}

\author[0000-0001-8587-9285]{Lucas Hunt}
\affiliation{National Radio Astronomy Observatory, P.O. Box O, Socorro, NM 87801, USA}

\begin{abstract}

The COSMOS HI Large Extragalactic Survey (CHILES) Continuum \& Polarization (CHILES Con Pol) survey is an ultra-deep continuum imaging study of the COSMOS field conducted using the Karl G. Jansky Very Large Array.  We obtained 1000 hours of L-band ($\lambda=20$ cm) observations across four spectral windows (1.063--1.831~GHz) on a single pointing and produced a confusion limited image with an apparent RMS noise of 1.67~\ujy\ beam$^{-1}$ with a synthesized beam of 5\farcs$5\times$5\farcs0. This paper reports a 1.4~GHz radio continuum source catalog containing 1678 sources detected above 7$\sigma$ (flux densities greater than 11.7~\ujy), identified using two independent source extraction programs applied to the Stokes {\it I} image. Resolved sources dominate at flux density $S_{\rm 1.4GHz} \ge 42$ \ujy. Radio spectral index for each source was derived using a power-law fit across the four spectral windows, and we found that a robust spectral index measurement requires a total signal-to-noise ratio of at least 20. Comparisons with previous 1.4~GHz radio continuum surveys show good overall consistency, but evidence for a high degree of catalog incompleteness and the effects of source confusion are evident for some of the earlier studies.

\end{abstract}

\keywords{Surveys (1671) --- Radio continuum emission (1340) --- Extragalactic radio sources (508)}

\section{Introduction} \label{sec:intro}

The evolution of galaxies over cosmic time has been a pivotal topic in modern astrophysics. The accumulation of stellar mass is modulated by star formation and its cessation, which can transpire through a diverse range of mechanisms, including Active Galactic Nuclei (AGN) feedback, gas depletion, and galaxy merging \citep{Heckman2014}. Elucidating the interplay between star formation and quenching is essential for achieving a comprehensive understanding of galaxy evolution across cosmic time and for comprehending the assortment of galaxy types that are observable in the present-day universe.

Deep radio continuum observation has emerged as an invaluable tool in the study of galaxy evolution \citep[see][ for a review]{Murphy2018}. Radio continuum serves as a dust-unbiased tracer for star formation and AGN activities, as it is dominated by synchrotron radiation emitted by cosmic rays that are accelerated by supernova remnants and AGN \citep{Condon1992}. Moreover, these observations are immune to dust obscuration, making them ideal for penetrating column densities that exceed $\rm N_{H} > 10^{24}$cm$^{-2}$ (corresponding to the optical extinction at V band $\rm A_{V} \gg 100$). Early studies of faint radio source have shown that radio sources fainter than 1~mJy at 1.4 GHz are a mixture of star-forming galaxies (SFG) and AGN, while AGNs dominate the sources at the higher flux density \citep[see the review by ][ and references therein]{deZotti2010}. Additionally, radio spectral energy distribution (SED) offers useful insight into the physical characteristics of the emitting sources: a steep spectral slope emanating from supernova remnants and a flat/inverted slope from an AGN core \citep{Condon1992}. Radio power also serves as a useful selection criterion for distinguishing between radio-loud and radio-quiet AGN, with radio power exceeding $P > 10^{25}$~W Hz$^{-1}$ typically indicating radio-loud AGN activity \citep{Miller1990}. Finally, a comparison of radio power with infrared (IR) luminosity can reveal even weak AGN activity, indicated by excess radio emission \citep{Yun2001, Bell2003}.

In recent years, sensitive radio interferometers, such as the Karl G. Jansky Very Large Array (VLA)\footnote{The National Radio Astronomy Observatory is a facility of the National Science Foundation operated under cooperative agreement by Associated Universities, Inc.}, MeerKAT, the Low Frequency Array (LOFAR), and the upgraded Giant Metrewave Radio Telescope (uGMRT), have achieved remarkable sensitivities in the \ujy\ range at the low-frequency radio continuum observations. For example, the COSMOS 2~deg$^{2}$ field was observed at 3~GHz using the VLA, resulting in a synthesized beam size of 0\farcs75 and a sensitivity of 2.3~$\mu$Jy~beam$^{-1}$ \citep{Smolcic2017a}. Another VLA study of the COSMOS field at both 3 and 10~GHz yielded synthesized beam sizes of 2\farcs0 and the root mean square (RMS) noise values of 0.53 and 0.41~$\mu$Jy beam$^{-1}$, respectively \citep[COSMOS-XS, ][]{vanderVlugt2021}. Recently, the DEEP2 field was observed at 1.28~GHz with MeerKAT, resulting in a beam size of 7\farcs6 and a thermal noise of 0.55~$\mu$Jy beam$^{-1}$ \citep{Mauch2020}. LOFAR and GMRT provide radio continuum observations at meter wavelengths. LOFAR Two-metre Sky Survey observed the deep fields of Bo\"{o}tes, Lockman Hole, and ELAIS-N1 fields at 150~MHz \citep{Tasse2021, Sabater2021}. Their synthesized beam size was 6\arcsec$\times$6\arcsec\, and the RMS noise values at the central regions were 32, 22, and 17~$\mu$Jy beam$^{-1}$ for Bo\"{o}tes, Lockman Hole, and ELAIS-N1 fields, respectively. GMRT observed the ELAIS-N1 field at 610~MHz with an RMS noise of 7.1~$\mu$Jy beam$^{-1}$ and a synthesized beam size of 6\arcsec$\times$6\arcsec\ \citep{Ocran2020}. Furthermore, \citet{Bera2018} reported an RMS noise of 21-39~$\mu$Jy beam$^{-1}$ for the DEEP2 field, with synthesized beam sizes ranging from 4\farcs7$\times$3\farcs9 to 6\farcs1$\times$4\farcs4 for four different sub-fields. Finally, \citet{Sinha2022} used uGMRT to observe the ELAIS-N1 field at 400~MHz, resulting in an RMS noise of 15~$\mu$Jy beam$^{-1}$ and a synthesized beam size of 4\farcs6$\times$4\farcs3.

More recently, the MeerKAT International GHz Tiered Extragalactic Exploration (MIGHTEE) conducted observations on the COSMOS and XMM-LSS fields, resulting in thermal noise of 1.7 and 1.5 $\mu$Jy beam$^{-1}$ and beam sizes of 8\farcs6 and 8\farcs2 at 1.4~GHz, respectively \citep{Heywood2022}. The wealth of sensitive high resolution data available now has ushered in the era of using these radio continuum data to provide independent constraints on key physical quantities such as star formation rates and AGN activities to the observational studies of galaxy evolution.

In this paper, we present the source catalogs from the COSMOS HI Large Extragalactic Survey (CHILES) Continuum Polarization (CHILES Con Pol or CCP, P.I. C. Hales) and delve into the discussions of their properties. The CHILES Con Pol survey was a commensal survey of the CHILES field, where the CHILES is the deep survey for detecting \HI\ 21\,cm emission directly from galaxies up to a redshift of 0.5 with a total integration time of 1000~hours. Four spectral windows (SPWs) were configured to get data in full polarization mode, widely separated in frequency and avoiding the worst radio frequency interference (Luber et al. in press, Paper 1 hereafter). The CHILES Con Pol data reduction and catalog generation are briefly described in \S~2.  The multi-wavelength data and redshifts of the CHILES Con Pol sources are described in \S~3, and the derived source properties including source size and spectral index distribution are described in \S~4. Throughout this study, we adopt the cosmological parameters from the WMAP7 in astropy, where $\Omega_{\rm M}=0.272$, $\Omega_{\rm \Lambda}=0.728$, and $H_{\rm 0}= 70.4$~km s$^{-1}$ Mpc$^{-1}$ \citep{Komatsu2011}.

\section{Observation \& Catalog Generations} \label{sec:obs}

\subsection{VLA observations}

The CHILES Con Pol observations employed four spectral windows (SPWs) chosen to avoid the range of frequencies with known strong radio frequency interference (RFI).  The central frequencies of the four chosen SPWs are 1.063, 1.447, 1.703, and 1.831~GHz. Each SPW featured a bandwidth of 128~MHz and 64 channels with a channel width of 2~MHz and full polarization products. Details of the experimental design as well as data reduction and imaging procedures are described in Paper 1. The final image, constructed using Briggs weighting \citep{Briggs1995} with a robust value of 0.5, 1024 wplanes, and multiterm (\texttt{nterms=4}) multifrequency synthesis imaging, has a synthesized beam size of 5\farcs5$\times$5\farcs0 (FWHM), an effective frequency of 1.447~GHz, and an RMS noise of 1.09 $\mu$Jy beam$^{-1}$, measured at $\geq 40$ arcminutes from the phase center in the Stokes {\it I} image. The RMS noise measured in the central 3\arcmin$\times$3\arcmin region was 1.67$\mu$Jy beam$^{-1}$. The higher noise in the pointing center is attributed to source confusion and imaging artifacts (see Paper 1 for a more comprehensive discussion). 

To facilitate the analysis of radio spectral indices, we produced separate continuum images for each SPW (see Paper 1 for further details). All four SPW images were smoothed to match the synthesized beam size of the lowest frequency SPW image, measuring 6\farcs5$\times$5\farcs5.

\subsection{Catalog Generations}

We generated two catalogs: one derived from the main continuum image and the second catalog from the SPW images where the spectral index could be determined. In the following subsections, we describe the methods used to extract sources from the continuum images.

\subsubsection{Main Continuum Source Catalog}

Radio interferometric observations with long integration time are susceptible to confusion noise from low flux density sources that fall below the detection limit, potentially biasing flux density measurements. Additionally, sources that are not fully deconvolved below the CLEAN threshold produce sidelobes, which appears as confusion-like noise in the image. The cumulative effect of their residual point spread function (PSF) leads to a measurable ``negative bowl'', and its characterization and correction are described in detail in Paper I.

In this study, we aim to accurately measure the flux densities of individual radio sources, which is a crucial step in performing source extraction, and we employed two different packages, namely Blobcat \citep{Hales2012} and Python Blob Detector and Source Finder \citep[PyBDSF, ][]{Mohan2015}. Both Blobcat and PyBDSF employ flood-fill algorithm to detect blobs exceeding a given signal-to-noise ratio (S/N) and isolate those surpassing a specified flooding S/N. For integrated flux density measurements, Blobcat integrates pixel values within the detected source region without relying on parabolic fitting, which is used solely for position determination \citep{Hales2012}. In contrast, PyBDSF implements 2D Gaussian fitting to estimate both peak and integrated flux densities \citep{Mohan2015}, potentially introducing systemic biases and uncertainties, particularly when measuring flux densities of extended sources. Given these methodological differences, we adopted the Blobcat-generated catalog as our primary source list.

We configured Blobcat with a detection S/N threshold of 5 and a flooding S/N threshold of 2.5, utilizing both signal and RMS images as inputs. The RMS and mean images were generated using the {\it AIPS} task RMSD, employing a 150-pixel radius centered on each pixel of interest, with signals above 3$\sigma$ excluded from RMS calculation. For comparison, PyBDSF was executed with equivalent parameters: a 5$\sigma$ detection threshold and a 2.5$\sigma$ island threshold, also incorporating RMS and mean images.

\begin{figure*}[h]
    \centering
    \includegraphics[width=7.1in]{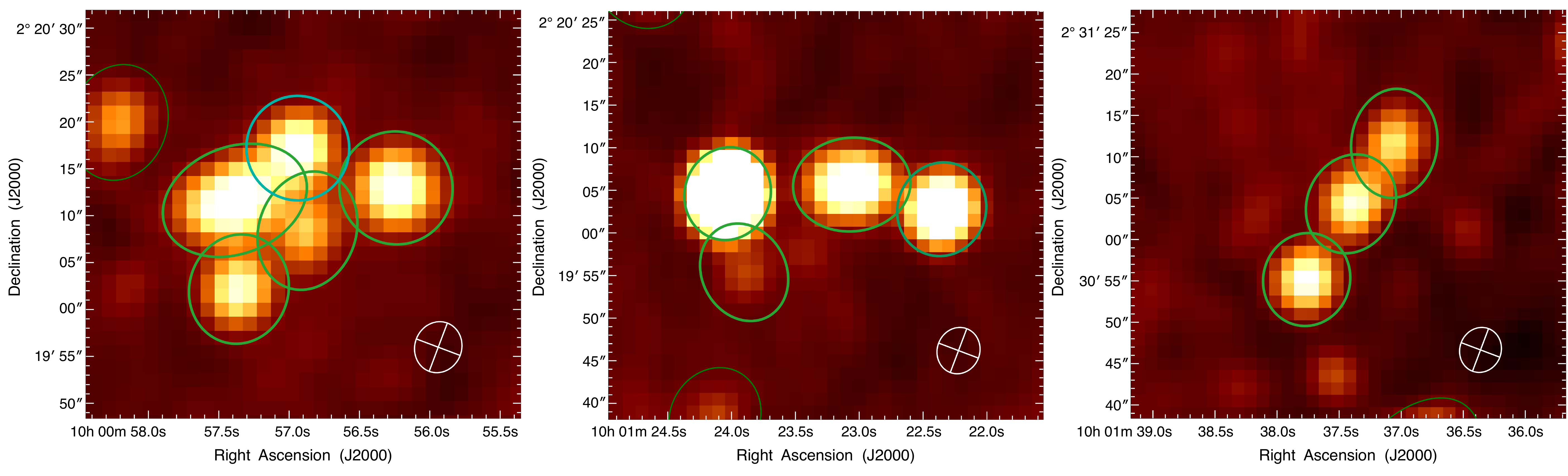}
    \caption{Three examples of source deblending performed by PyBDSF. In each case, Blobcat had difficulty fully separating the five (left), four (middle), and three (right) sources in these examples. The white ellipse in lower right corner of each panel represents the synthesized beam size of the CHILES Con Pol. 
    }
    \label{fig:blended}
\end{figure*}

A major shortcoming of Blobcat is that it is not able to deblend overlapping sources (Hales et al. 2012). To address this constraint, we complemented our analysis with PyBDSF for deblending confused sources, as illustrated in Figure~\ref{fig:blended}. By cross-referencing both catalogs, we classified sources as blended when multiple PyBDSF detections fell within a single Blobcat source region, and as single when there was a one-to-one correspondence between PyBDSF and Blobcat detections. For blended sources, we adopted positions and flux densities from PyBDSF, while for single sources, we utilized Blobcat measurements. While flux density measurements from both software packages demonstrated strong concordance for single sources, we preferred Blobcat's model-independent photometry to eliminate potential systematic biases that could arise from morphological assumptions. A total of 322 blended sources are decomposed into 556 sources by PyBDSF, along with 1958 single sources. Eight resolved sources are detected exclusively by Blobcat (“Blobcat only”), while PyBDSF uniquely identified one unresolved source (``PyBDSF only’’) source. The ``Blobcat only’’ sources exhibit S/N ranging from 7.04 to 14.24 (specifically, 7.04, 7.08, 7.18, 7.65, 7.71, 8.02, 10.37, and 14.24), whereas the single ``PyBDSF only’’ source presents an S/N of 7.52. These statistics underscore Blobcat's sensitivity to extended sources that PyBDSF may overlook, while PyBDSF proves more adept at identifying point sources evading detection by Blobcat. The final catalog comprises of 2523 and 1678 sources exceeding 7$\sigma$ in peak flux density within 20\% and 50\% of the PB, respectively. In the remainder of this paper, we consider only the catalog of 1,678 radio sources exceeding 7$\sigma$ within the 50\% PB region.

\subsubsection{Determining Spectral Index from Individual SPW Images}

Deriving radio spectral index of each CHILES Con Pol source requires identifying matching source in each of the individual SPW images and modeling their SED in a power-law form.  As a first step, the CHILES Con Pol continuum images of each SPW were generated using the same procedure as that of the main CHILES Con Pol image discussed above. To ensure consistency, we adjusted the synthesized beam sizes of the SPW images to match that of the lowest frequency SPW 0 (1.063~GHz) image, with $\theta =$ 6\farcs5$\times$5\farcs5. Source extraction for each SPW image was done using a lower S/N threshold of 3.5, given that each SPW image has a local RMS noise approximately twice that of the main image.  The ``negative bowl" correction was also carried out for each SPW image before running source extraction.

Sources with flux density near the catalog detection limit suffer from the well-known problem of flux boosting and greater position uncertainty, and this leads to a significant uncertainty in the derived spectral index. To mitigate this problem, we restricted the offsets of source peak positions to within 2 pixels (3\arcsec). Our analysis yielded radio spectral indices for 1614 sources (96.2\%), determined through power-law fitting of flux densities across two or more SPWs. Given the relatively narrow frequency range covered by CHILES Con Pol, it is reasonable to assume that the flux densities across the SPWs follow a power-law. The fitting was conducted using the \pkg{lm} function in \textit{R} \citep{rcite}, with weights corresponding to 1/($\epsilon_{peak}^{2}$) for the uncertainty of the peak flux density, $\epsilon_{peak}$. The large uncertainty in the spectral index measurements is primarily due to the low S/N and the limited frequency coverage of the four SPWs. 

The CHILES Con Pol source catalog (Table~\ref{tab:catalog}) include: (1) source ID, (2) right ascension as [h m s] (RA); (3) uncertainty in RA as [\arcsec] (e\_RA); (4) Declination as [\degree\ \arcmin\ \arcsec] (DEC); (5) uncertainty in DEC as [\arcsec] (e\_DEC); (6) 1.4~GHz peak flux density in Jy (S$_{\rm peak}$); (7) uncertainty of 1.4~GHz peak flux density in Jy (e\_S$_{\rm peak}$); (8) 1.4~GHz integrated flux density in Jy (S$_{\rm int}$); (9) uncertainty of 1.4~GHz integrated flux density in Jy (e\_S$_{\rm int}$); (10) spectral index (SI); and (11) uncertainty of spectral index (e\_SI). The full $7\sigma$ CHILES Con Pol catalog is available in digital form online.

\centerwidetable
\begin{table*}
\caption{CHILES Con Pol radio continuum source catalog}
\label{tab:catalog}
\scriptsize
\begin{tabular}{rcccccccccc}
\hline \hline
(1) & (2) & (3) & (4) & (5) & (6) & (7) & (8) & (9) & (10) & (11) \\
ID & RA & e\_RA & DEC & e\_DEC & S$_{\rm peak}$ & e\_S$_{\rm peak}$ & S$_{\rm int}$ & e\_S$_{\rm int}$ & SI & e\_SI \\
 & (h m s) & (\arcsec) & (\degree\ \arcmin\ \arcsec) & (\arcsec) & (Jy) & (Jy) & (Jy) & (Jy) & & \\
\hline
   1 & 10 00 26.938 & 0.131 & 02 22 31.368 & 0.141 & 1.137e-4 & 7.373e-6 & 1.881e-4 & 1.026e-5 & -0.566 &  0.272 \\
   2 & 10 00 26.945 & 0.347 & 02 23 14.676 & 0.374 & 4.181e-5 & 4.605e-6 & 4.181e-5 & 4.605e-6 & -0.650 & 0.470 \\
   3 & 10 00 27.437 & 0.027 & 02 20 57.048 & 0.029 & 6.895e-4 & 3.739e-5 & 1.089e-2 & 5.448e-4 & -999 & -999 \\
   4 & 10 00 27.631 & 0.336 & 02 19 16.572 & 0.360 & 4.269e-5 & 4.590e-6 & 4.269e-5 & 4.590e-6 & -0.550 &  0.627 \\
   5 & 10 00 28.188 & 0.479 & 02 19 47.640 & 0.515 & 3.034e-5 & 4.352e-6 & 3.034e-5 & 4.352e-6 & -0.134 &  0.611 \\
   6 & 10 00 28.344 & 0.490 & 02 17 31.884 & 0.526 & 2.695e-5 & 3.925e-6 & 2.695e-5 & 3.925e-6 & -0.824 &  0.888 \\
   7 & 10 00 28.579 & 0.227 & 02 19 28.236 & 0.244 & 6.277e-5 & 5.200e-6 & 6.277e-5 & 5.200e-6 & -0.536 &  0.470 \\
   8 & 10 00 28.704 & 0.228 & 02 17 44.808 & 0.245 & 5.728e-5 & 4.750e-6 & 5.728e-5 & 4.750e-6 & -0.389 &  0.458 \\
   9 & 10 00 29.234 & 0.508 & 02 23 40.272 & 0.547 & 2.530e-5 & 3.819e-6 & 2.530e-5 & 3.819e-6 & -999 & -999 \\
  10 & 10 00 29.462 & 0.511 & 02 25 36.300 & 0.547 & 2.488e-5 & 3.764e-6 & 3.144e-5 & 3.853e-6 & -0.867 &  1.033 \\
\hline
\end{tabular}
\tablecomments{The first 10 sources from the full CHILES Con Pol $7\sigma$ catalog are shown as an example. Radio sources lacking a spectral index measurement are indicated with a ``-999'' in  columns (10) and (11). The complete catalog is available in the machine-readable format online.}
\end{table*}

\subsection{Completeness \label{sec:completeness}}

Completeness in source extraction refers to the extent to which our method recovers sources at different flux density level. A Monte-Carlo simulation is employed to measure completeness by generating images with a set of test sources inserted and assessing how many of these test sources are successfully recovered. Test sources consist of 2D Gaussians of the synthesized beam size and beam position angle, where $\sigma=\theta_{\rm FWHM}/(2\sqrt{2 \; \rm ln (2)})$ for the beam size, $\theta_{\rm FWHM} =$ 5.5\arcsec $\times$ 5.0\arcsec. A total of 300 test sources with flux density corresponding to S/N values of 7, 7.5, 8, 9, 10, 15, 20, 30, 40, 50, 75, 100, 150, 200, 300, 400, and 500 are added to the image for this completeness test. This process is repeated five times for each S/N value.

\begin{figure}[!h]
    \centering
    \includegraphics[width=4.1in]{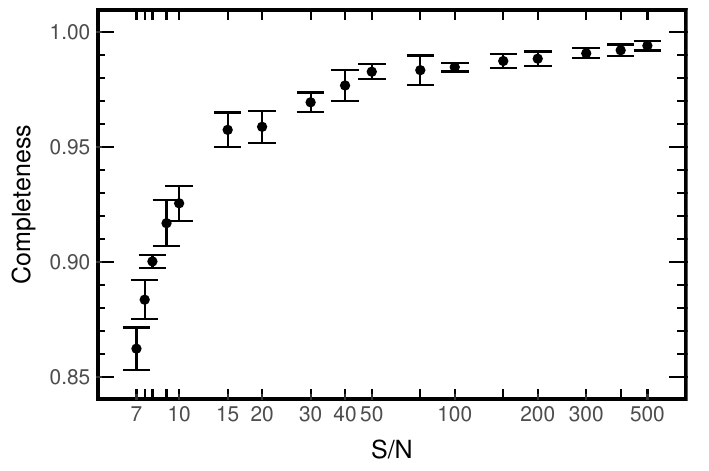}
    \caption{Catalog completeness as a function of an input S/N. The completeness is measured through the Monte-Carlo simulations and the error bar indicates 1$\sigma$ uncertainty. Our catalog is complete 86\% at 7$\sigma$ and 96\% at S/N$=$15.}
    \label{fig:completeness}
\end{figure}

Source extraction in this simulation is conducted solely with PyBDSF, since the inserted test sources are  point sources with an FWHM equivalent to the beam size. Test sources are generally identified at their input positions. However, some of the test sources are confused or blended with existing sources in the CHILES Con Pol image, and their recovered source flux density can differ from the input flux density. To account for this effect, we consider a test source ``recovered" if their derived flux density is within 50\% of the input flux. The resulting estimates of the catalog completeness is shown as a function of the S/N in Figure~\ref{fig:completeness}. The catalog completeness reaches 96\% at S/N$\sim$15 and nearly 99\% at S/N$>$150.

\section{Multi-wavelength Data}

There is a wealth of photometric datasets available in the COSMOS field, spanning the wavelengths from X-ray through millimeter. This study leverages this rich multi-wavelength data  to estimate the radio luminosity of radio continuum sources identified by the CHILES Con Pol survey. Specifically, we utilize data from the \textit{Spitzer Space Telescope} (\textit{Spitzer}) Infrared Array Camera (IRAC), the spectroscopic redshift compilations from G10/COSMOS \citep{Driver2018}, the Deep Imaging Multi-Object Spectrograph 10k catalog \citep[Deimos 10k, ][]{Hasinger2018}, and the photometric redshift catalog from the compilation catalogs of optical-to-near infrared observations, COSMOS 2020-R1 \citep{Weaver2022}. This multi-wavelength data is also employed in a forthcoming paper, which will explore the origins of synchrotron emission through SED analysis and AGN identification (Gim et al. in prep.)

\subsection{Spitzer IRAC Data}

Data from the \textit{Spitzer}/IRAC survey by \citet[][S-COSMOS]{Sanders2007}\footnote{Data was downloaded at https://irsa.ipac.caltech.edu/data/SPITZER/S-COSMOS/} are particularly important for identifying multi-wavelength counterparts for the CHILES Con Pol radio sources and determining their spectroscopic and photometric redshifts. The S-COSMOS program dedicated 166 hours of observations to a deep survey of the entire 2 deg$^{2}$ of the COSMOS field, achieving a sensitivity of 0.18 $\mu$Jy at 3.6 $\mu$m. 
The high angular resolution (1\farcs8) and superb sensitivity of \textit{Spitzer} IRAC data are both critically important for identifying the counterparts for the radio sources using the likelihood ratio method \citep{sutherland92} and the nearest neighbor matching.  This analysis led to a successful identification of total of 1666 (99.3\%) IRAC counterparts to the radio continuum sources.

\subsection{UV-NIR Data \label{sec:cosmos2020}}

Identifying a correct optical/UV galaxy and associated ultraviolet (UV) to near-infrared (NIR) photometry is a critical step in determining the redshift and radio luminosity of each radio source.  We utilized the COSMOS 2020-R1 catalog \citep{Weaver2022} to access its compilation of rich multi-wavelength datasets, utilizing the IRAC-radio match discussed above.  The COSMOS 2020-R1 catalog integrates data collected from a variety of telescopes and instruments, spanning the UV to NIR wavelength ranges. Specifically, the UV data were sourced from the GALEX in Far-UV (FUV) at 150~nm and Near-UV (NUV) at 230~nm. Optical data were acquired through observations in broad bands such as u, u*, g, r, i, z, y by the Subaru/Hyper Suprime-Cam, as well as B, g$+$, V, r$+$, i$+$, z$+$, z$++$ by Subaru/Suprime-Cam. Additionally, two narrow bands centered on 711~nm and 816~nm from Subaru/Suprime-Cam, along with ACS F814W from the \textit{Hubble Space Telescope}, were included in the catalog. The NIR data encompassed observations made in the Y, J, H, and Ks broad bands, as well as NB118 narrow bands, all of which were acquired using VISTA/VIRCAM. Furthermore, the \textit{Sptizer} contributed data in the form of IRAC observations at 3.6, 4.5, 5.8, and 8.0~$\mu$m. This comprehensive dataset yields photometric measurements conducted by two independent methods: aperture photometry using PSF-matched images (CLASSIC catalog) and profile-fitting photometry employing the Tractor (FARMER catalog). 

For identifying counterparts for radio-matched IRAC sources, we adopt the CLASSIC catalog with 3\arcsec\ aperture photometry. We have identified 1559 (93.6\%) UV-NIR counterparts to the radio-matched IRAC 3.6$\mu$m sources using the nearest neighbor matching and a 2\arcsec\ search radius.

\subsection{Redshifts}

\subsubsection{Spectroscopic Redshifts \label{sec:specz}}

Spectroscopic redshifts (spec-z) for radio sources are compiled from two primary sources: the G10/COSMOS v0.5 catalog \citep{Driver2018} and the Deimos 10k catalog \citep{Hasinger2018}. The G10/COSMOS catalog provides the ``best'' redshift using a parameter $z_{\rm use}$: $z_{\rm use} \leq 3$ for a spec-z and $z_{\rm use} = 4$ or 10 for photometric redshifts (photo-z). 
We identified 1404 best redshifts from the G10/COSMOS catalog, of which 681 are spectroscopic redshifts. Additionally, we found 109 spec-z within the Deimos 10k catalog. In total, 790 (47.3\%) CHILES Con Pol sources have a spectroscopic redshift.

\subsubsection{Photometric Redshifts \label{sec:photz}}

We utilize the phot-z compiled in COSMOS 2020-R1 \citep{Weaver2022}, as discussed in Section~\ref{sec:cosmos2020}. The phot-z in COSMOS 2020-R1 are derived using the \pkg{LePhare} \citep{Arnouts2002, Ilbert2006} and \pkg{EAZY} \citep{Brammer2008} codes, and two estimates generally agree quite well.
From the crossmatching described in Section~\ref{sec:cosmos2020}, we identified 1559 counterparts within the COSMOS 2020-R1, and a photo-z is available for  1523 counterparts. We adopt these photo-z for all CHILES Con Pol radio sources without spec-z.

\begin{figure}[!h]
    \centering
    \includegraphics[width=4.3in]{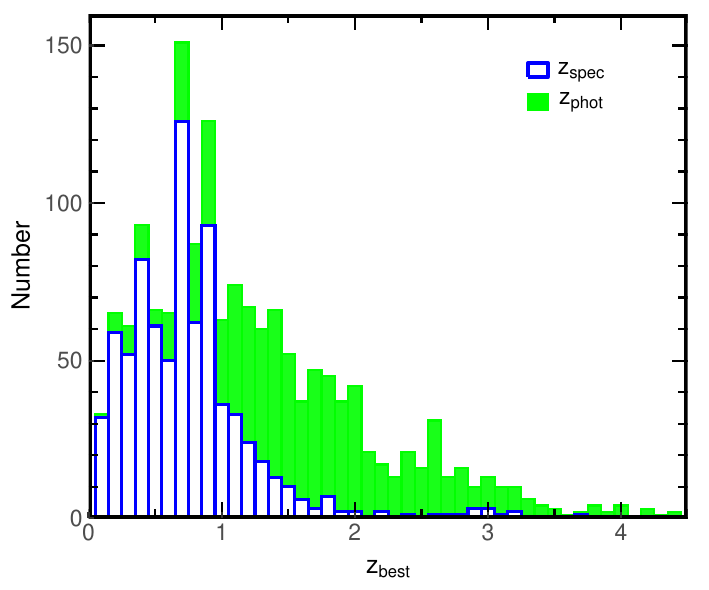}
    \caption{The histogram of redshifts is presented with a bin size of 0.1, where $z_{spec}$ is shown with the blue line and $z_{phot}$ is with the green filled histogram. Majority (51.0\%) are located within $z<1$.}
    \label{fig:z_distr}
\end{figure}

\subsubsection{Combined Redshifts \label{sec:allz}}

By combining 790 spec-z and 797 phot-z, a redshift is available for a total of 1587 sources (95.3\% of the full sample). As shown in Figure~\ref{fig:z_distr}, the majority (51.0\%) of CHILES Con Pol sources have a redshift $z < 1$, with a mean redshift of $z=1.21$ and a median of $z=0.98$. Almost all spec-z sources, shown by the empty histogram, are within the redshift range below 1.5. Our redshift distributions are consistent with the results of VLA-COSMOS 3~GHz Large Project \citep{Smolcic2017b}, which also showed a decline in the number of $z_{spec}$ beyond $z \sim 1$. A narrow peak in the redshift distribution ($\sim 30$ galaxies) seen near $z\approx 2.6$ might be tracing the radio sources associated with the proto-cluster of star forming galaxies previously reported by \citet{Casey2015}.

\section{Properties of the radio sources}

\subsection{Resolved vs Unresolved \label{sec:size}}

The determination of whether a source is resolved or unresolved in astronomical observations is commonly based on the ratio of integrated flux density to peak flux density. Specifically, the inclusion of extended emission from a resolved source results in a larger integrated flux density relative to the peak flux density, whereas the presence of noise leads to a smaller integrated flux density in an unresolved source \citep{Bondi2003, Smolcic2007}. To this end, we calculated the ratio of integrated flux density to peak flux density, $Ratio = S_{int} /  S_{peak}$ and its associated uncertainty $\sigma_{Ratio}$. Subsequently, we defined resolved sources as those with $Ratio-\sigma_{Ratio} > 1$, and unresolved sources as those that do not meet this criterion. 

\begin{figure*}[ht]
    \centering
    \includegraphics[width=7.1in]{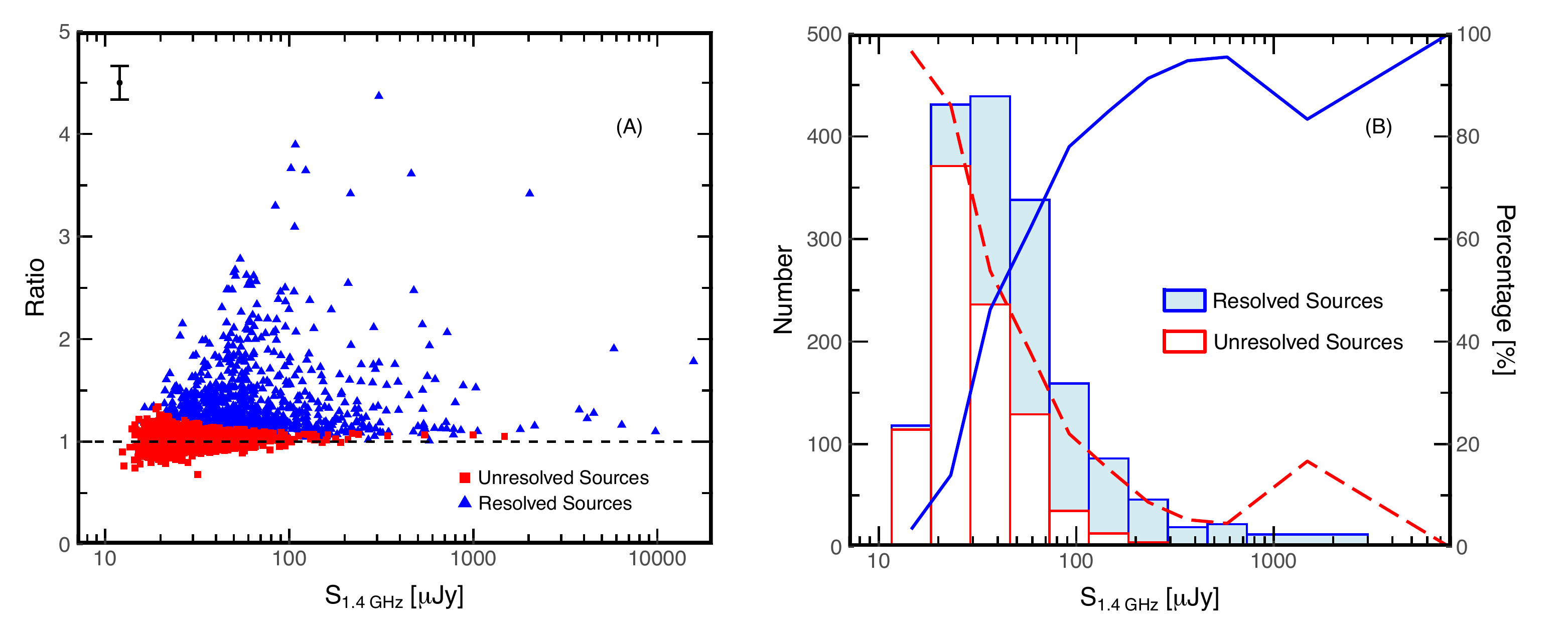}
    \caption{The ratio of integrated flux density to peak flux density of the sources detected in the CHILES Con Pol survey (left panel) and histograms and fractions of unresolved and resolved sources (right panel) as a function of flux density. (A) The resolved sources based on the integrated to peak flux ratio are shown with blue filled triangles, while the unresolved sources are shown with red filled squares. The horizontal black dashed line indicates the unity ratio. Typical (median) uncertainty in this ratio ($\sigma=0.18$) and 10\% flux density are shown on the upper left of the plot. (B) Histograms of unresolved sources (blank red bars) and resolved sources (filled light blue bars) are shown  with the bin size of $\rm \Delta log \; S_{1.4GHz} = 0.2$. The percentage of unresolved sources (red dashed line) and resolved sources (blue solid line) are also shown. }
    \label{fig:resolved}
\end{figure*}

The ratio of integrated flux density to peak flux density ($Ratio$) as a function of the integrated flux density (panel A) and histograms and fractions of unresolved/resolved sources as a function of flux density (panel B) are shown in Figure~\ref{fig:resolved}. The left panel (Panel A) shows that the radio sources in our sample tend to be more concentrated near a ratio of 1 (black dashed line), suggesting that the majority of our radio sources are unresolved or partially resolved at this resolution. Using a formal definition of unresolved/resolved sources based on this ratio, we identify a total of 772 (46\%) resolved sources and 906 (54\%) unresolved sources in our dataset.

Histograms of resolved (light blue bars) and unresolved (blank red bars) sources are shown as a function of flux density with a bin size of $\rm  \Delta log \; S_{1.4GHz} = 0.2$ on the left panel of Figure~\ref{fig:resolved}. The fraction of resolved (blue solid line) and unresolved (red dashed line) plotted on the same panel clearly show that the fraction of resolved sources increases as flux density. Unresolved sources account for 96.6\% at the faintest flux density bin of 11.7$-$18.4~\ujy\ and 4.5\% at the flux density bin of 462$-$732~\ujy. Flux density bins above 1 mJy have small number of sources and thus are subject to a large statistical uncertainty. Resolved sources dominate at flux density $S_{1.4GHz} \ge 42$~\ujy, in accordance with earlier findings in the deep SWIRE field \citep{Owen2008} and Great Observatories Origins Deep Survey-North field \citep{Owen2018}. Our source identification methods have a systematic bias against finding extended sources with peak brightness below the detection threshold.  The significance of this potential bias is difficult to estimate, but this bias may not be significant since compact sources dominate the fraction up to 40 $\mu$Jy, well above the catalog detection limit. These statistics, along with the uncertainties estimated using Poisson statistics, are summarized in Table~\ref{tab:fractions}. 

\begin{table}[h]
    \centering
    \caption{Fractions of unresolved and resolved sources}
    \begin{tabular}{c|c|c}
    \hline
        flux density (\ujy)  & unresolved (\%) & resolved (\%) \\
    \hline
      11.6 -- 18.4 & 96.6$\pm$9.0 & 3.4$\pm$1.7 \\
      18.4 -- 29.1 & 86.1$\pm$4.5 & 13.9$\pm$1.8 \\
      29.1 -- 46.2 & 53.8$\pm$3.5 & 46.2$\pm$3.2 \\
      46.2 -- 73.2 & 38.2$\pm$3.4 & 61.8$\pm$4.3 \\
      73.2 -- 116.0 & 22.0$\pm$3.7 & 78.0$\pm$7.0 \\
      116.0 -- 183.8 & 15.1$\pm$4.2 & 84.9$\pm$9.9 \\
      183.8 -- 291.4 & 8.7$\pm$4.3 & 91.3$\pm$14.1 \\
      291.4 -- 461.8 & 5.3$\pm$5.3 & 94.7$\pm$22.3 \\
      461.8 -- 731.9 & 4.5$\pm$4.5 & 95.5$\pm$20.8 \\
      731.9 -- 3000 & 16.7$\pm$11.8 & 83.3$\pm$26.4 \\
    \hline
    \end{tabular}
    \label{tab:fractions}
\end{table}

\subsection{Radio Spectral Index Distribution}
\subsubsection{Dependence on S/N}

The low-frequency radio spectral index can serve as a useful indicator of the mechanism responsible for the observed emission. The accuracy of this measurement is highly sensitive to the methodology employed, and the spectral index measurements are reliable only for sources detected with a sufficiently high S/N. The wide bandwidths of the VLA have facilitated in-band spectral index measurements through multi-term multi-frequency synthesis imaging in CASA \citep[e.g., ][]{Rau2016, Herrero-Illana2017, Kumar2023}. Nonetheless, experiments with the CHILES Con Pol frequency setup have demonstrated that in-band spectral index estimates are less accurate compared to those derived from a power-law fit to the flux densities measured at each SPW \citep{Gim15}. The unique SPW setup of the CHILES Con Pol survey, particularly the narrow frequency separation between SPW 3 and 4 ($\Delta \nu =128$~MHz), results in a flatter estimation of the in-band spectral index. Consequently, we estimate the spectral index of our sources by fitting a power-law model, $S \sim \nu^{\alpha}$, to the flux densities measured across the four SPW images.

\begin{figure*}[!ht]
    \centering
    \includegraphics[width=7.1in]{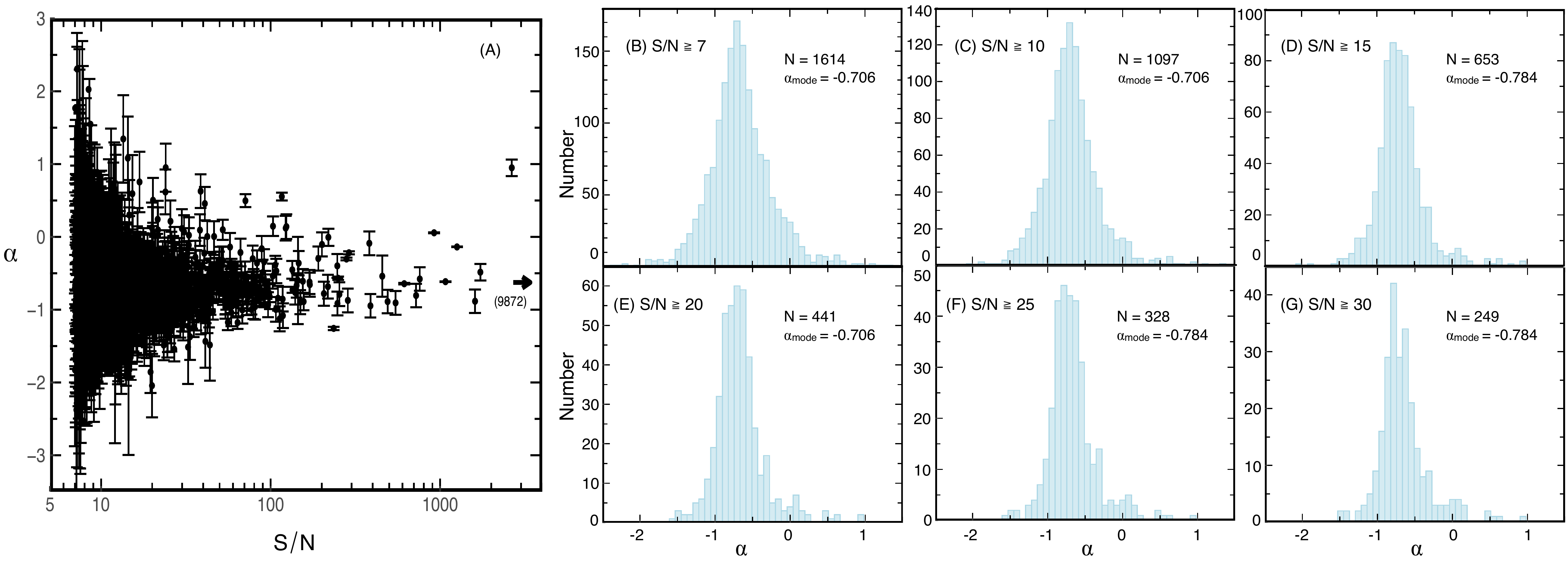}
    \caption{The radio spectral index distribution of the CHILES Con Pol radio continuum sources. The peak of the histogram is placed on the $\alpha=-0.725$, implying that the majority of $\mu$Jy radio sources are dominated by star formation. Panels (B)-(G) display histograms of the spectral index along the S/N cut, e.g., 7, 10, 15, 20, 25, and 30.}
    \label{fig:si_distr}
\end{figure*}

A plot of the measured radio spectral indices as a function of S/N is shown on the left panel of Figure~\ref{fig:si_distr}.  The spectral index distribution is very broad with large uncertainties for sources with low S/N (S/N$\lesssim$20). In addition, significant uncertainties are found for the cases where the spectral index was estimated solely from the SPW 3 and 4 data, due to the narrow frequency separation between these SPWs (128~MHz). 

To assess the reliability of the radio spectral index measurements, we examine the spectral index distributions derived with varying S/N thresholds. The histograms of the radio spectral indices are shown in panels (B)-(G) of Figure~\ref{fig:si_distr} as a function of S/N. The bin size is chosen by the Freedman-Diaconis rule \citep{Freedman1981} with S/N$>$7 sources, $\Delta \alpha= 0.078$. These histograms show that sources with extreme spectral indices ($\alpha < -1.5$ and $\alpha > 1.0$) disappear at S/N$>$20 with the 95\% range of $[-1.25, 0.14]$, strongly indicating that these extreme spectral indices are the results of noisy measurements.  The spectral index distributions are similar and more consistent with each other at S/N threshold of 20 and larger. 

In panel (B), the mode of the distribution is located at $\alpha= -0.706$, consistent with the typical spectral index of optically thin synchrotron emission from star formation \citep{Condon1992, Magnelli2015, Smolcic2017a, Gim2019}. This observation suggests that the majority of $\mu$Jy sources in our sample are likely associated with star formation processes. As S/N increases, the widths of the distributions narrow, while the mode remains nearly the same. It is also apparent that the spectral index distribution is not symmetric, with a secondary peak near $\alpha\approx 0$, which is more obvious in the two highest S/N threshold cases in panels (F) \& (G). Synchrotron emission at radio frequency is composed of optically thin synchrotron radiation, self-absorbed synchrotron (synchrotron self-absorption), and aged synchrotron (synchrotron aging) components. Synchrotron self-absorption arises in optically thick regions with high brightness temperature $T_{b} \sim 10^{9}$~K \citep[e.g., AGN cores and jet bases, ][]{Blandford1979, Eckart1986}, exhibiting a maximum spectral index of $2.5$ \citep{Rees1967}. Aged synchrotron, in contrast, occurs when relativistic electrons with high energies in magnetic fields are preferentially depleted due to more efficient radiation compared to their lower energy counterparts over time, resulting in spectral steepening, typically characterized by spectral indices $\alpha \gtrsim -2$ \citep[e.g., in radio lobes and jets,][]{Carilli1991, Jamrozy2008, Orru2010}. The secondary peak near $\alpha\approx 0$ is likely associated with the known population of relatively rare ``flat spectrum" radio sources that are generally associated with cores of radio AGNs.

\subsubsection{Dependence on radio power}

\begin{figure}[!h]
    \centering
    \includegraphics[width=7in]{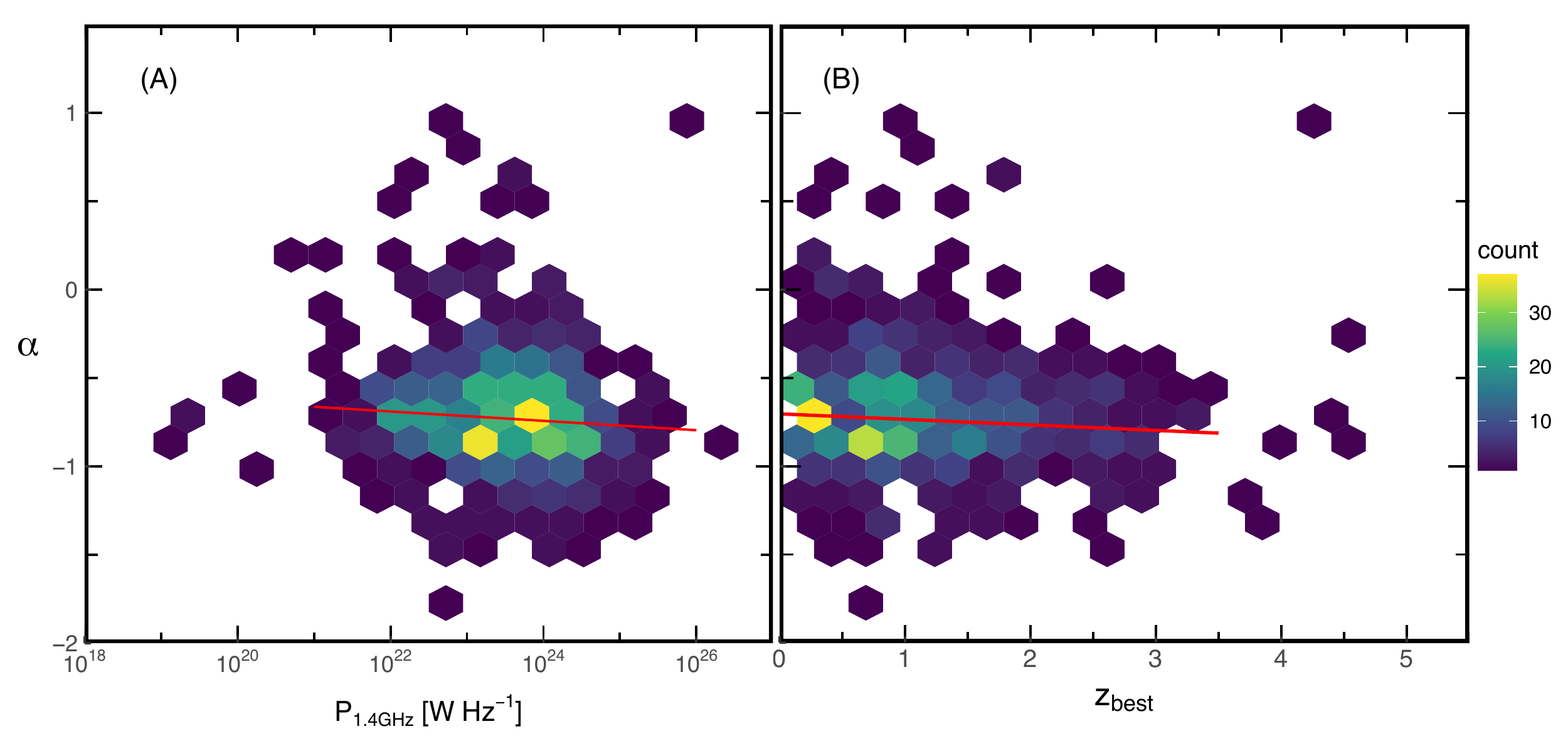}
    \caption{Hexbin plots (2D histograms) of radio spectral index as a function of 1.4~GHz power (panel A) and z$_{best}$ (panel B). The red line indicates the linear regression computed by a robust linear model (\pkg{lmRob} function) in R \citep{rcite}. }
    \label{fig:si_lum_z}
\end{figure}

We further explore the relationship between the spectral index and radio power or redshift to investigate the nature of the width of spectral index distribution. Previous studies have shown an anti-correlation between spectral index and radio power in bright Fanaroff–Riley II galaxies. At higher radio powers, spectral indices steepen due to the radiative losses occurring in the amplified magnetic fields of the hotspots within galaxies that have more powerful jets \citep{Laing1980, Blundell1999}. This trend has also been observed in sub-mJy sources, where core-dominated sources with flat spectra emerge \citep{Prandoni2006, Whittam2013}. In panel (A) of Figure~\ref{fig:si_lum_z}, we examine the correlation between radio spectral index and 1.4~GHz radio power using a hexbin plot. To ensure robust estimates, we limit our analysis to sources with $10^{21} \le P_{\rm 1.4GHz} \le 10^{26}$~W Hz$^{-1}$ and $z \le 3.5$, and restrict the sample to those with S/N$>$20 to minimize uncertainties in spectral index.

Using a robust linear model (\pkg{lmRob} function) and bootstrap resampling (\pkg{boot} function) in R \citep{rcite}, we estimate the slope and intercept of the relation, along with their associated uncertainties (summarized in Table~\ref{tab:si_lum_z}), yielding the model (red line in Figure~\ref{fig:si_lum_z})  $\alpha=-0.110 (^{+0.319}_{-0.324}) - 0.026 (\pm 0.014) \, \rm log_{10} \left(P_{1.4GHz} \right)$.The mean range of spectral index is from -0.655 to -0.785 for $10^{21} \le P_{\rm 1.4 GHz} \le 10^{26}$~W Hz$^{-1}$, which falls within the mean dispersion in spectral index, $0.23$. We note that our results are inconsistent with previous studies at higher frequency reporting the spectral steepening, such as -0.28 at $0.1 < S_{\rm 4.86 GHz} < 0.2$~mJy to -0.73 at $0.2 < S_{\rm 4.86 GHz} < 16.5$~mJy \citep{Ciliegi2003}, from -0.29 for $S_{\rm 5 GHz} \leq 4$~mJy to -0.62 for $S_{\rm 5 GHz} > 4$~mJy \citep{Prandoni2006}, and from -0.11 at $0.3 < S_{\rm 15.7 GHz} \leq 0.755$~mJy to -0.66 at $1.492 < S_{\rm 15.7 GHz} < 45.7$~mJy \citep{Whittam2013}.  
The absence of steepening in our data is expected, as our sample does not include rare sources with radio powers $\sim 10^{27-29}$~W Hz$^{-1}$ \citep{Blundell1999}, where steepening is the most pronounced.

A correlation between the radio spectral index and redshift has been reported previously, with steeper spectra commonly found in high redshift radio galaxies \citep{Tielens1979}. Even though the underlying cause of this correlation remains unclear, three possible explanations have been proposed: i) spectral steepening due to a concave spectrum, driven by synchrotron self-absorption at low frequencies and the balance between synchrotron and inverse Compton losses at high frequencies; ii) an indirect effect of $P_{\rm 1.4GHz} - \alpha$ correlation; and iii) increased ambient density at high redshifts \citep[see][ for a detailed review]{Miley2008}. We quantified the relationship between the radio spectral index and redshift using the same method employed for the $P_{\rm 1.4GHz}- \alpha$ correlation as shown in panel (B) of Figure~\ref{fig:si_lum_z}. The best-fit model yields the relation of $\alpha=-0.684 (^{+0.030}_{-0.013}) - 0.041 (\pm 0.016) z_{best}$. 
The mean range of the spectral index is from -0.684 to -0.829 for $0 \le z \le 3.5$, which remains within the mean uncertainty of $0.23$, similar to the results from the $P_{\rm 1.4 GHz}-\alpha$ relation shown in Panel (A). 
Furthermore, the lack of significant evolution in radio spectral index within $0<z<3$ is consistent with the recent studies on the sub-mJy radio sources \citep{Magnelli2015, Calistro-Rivera2017, Retana-Montenegro2022}. Our findings suggest that the production of synchrotron radiation is not influenced by global redshift-dependent effects but is instead governed by local processes such as magnetic field \citep{schleicher13} and other cosmic ray energy loss mechanisms, including ionization, free-free emission, and inverse Compton losses \citep{lacki10a}.

Finally, our analysis of the trends seen in Figure~\ref{fig:si_lum_z} indicates no statistically significant steepening of the spectral index with increasing radio power or redshift, with the significance of the slopes corresponding to 2.2$\sigma$ for both radio power and redshift. Pearson's correlation test yields coefficients of -0.17 for the $\alpha-P_{\rm 1.4~GHz}$ relation and -0.15 for the $\alpha-z_{\rm best}$ relation, meaning that the weak tendencies for steeper spectral indices at higher radio powers and redshifts are statistically insignificant due to large uncertainties. Future observations with higher sensitivity and larger spectral baselines should provide a more accurate characterization of the spectral index distribution.

\begin{table}[h]
    \centering
    \caption{Parameters of robust linear models of radio spectral index versus P$_{\rm 1.4 GHz}$ and z$_{\rm best}$. }
    \begin{tabular}{c|c|c}
    \hline
        parameter & intercept & slope \\ 
    \hline
        $\alpha$ vs $\rm log_{10} \;(P_{1.4GHz}$) & -0.110$^{+0.319}_{-0.324}$ & -0.026$\pm 0.014$ \\
        $\alpha$ vs $z_{\rm best}$  & -0.684$^{+0.030}_{-0.013}$ & -0.041$\pm 0.016$ \\
    \hline
    \end{tabular}
    \label{tab:si_lum_z}
\end{table}

\subsection{1.4 GHz Radio Power Distributions}

\begin{figure}[!h]
    \centering
    \includegraphics[width=7.3in]{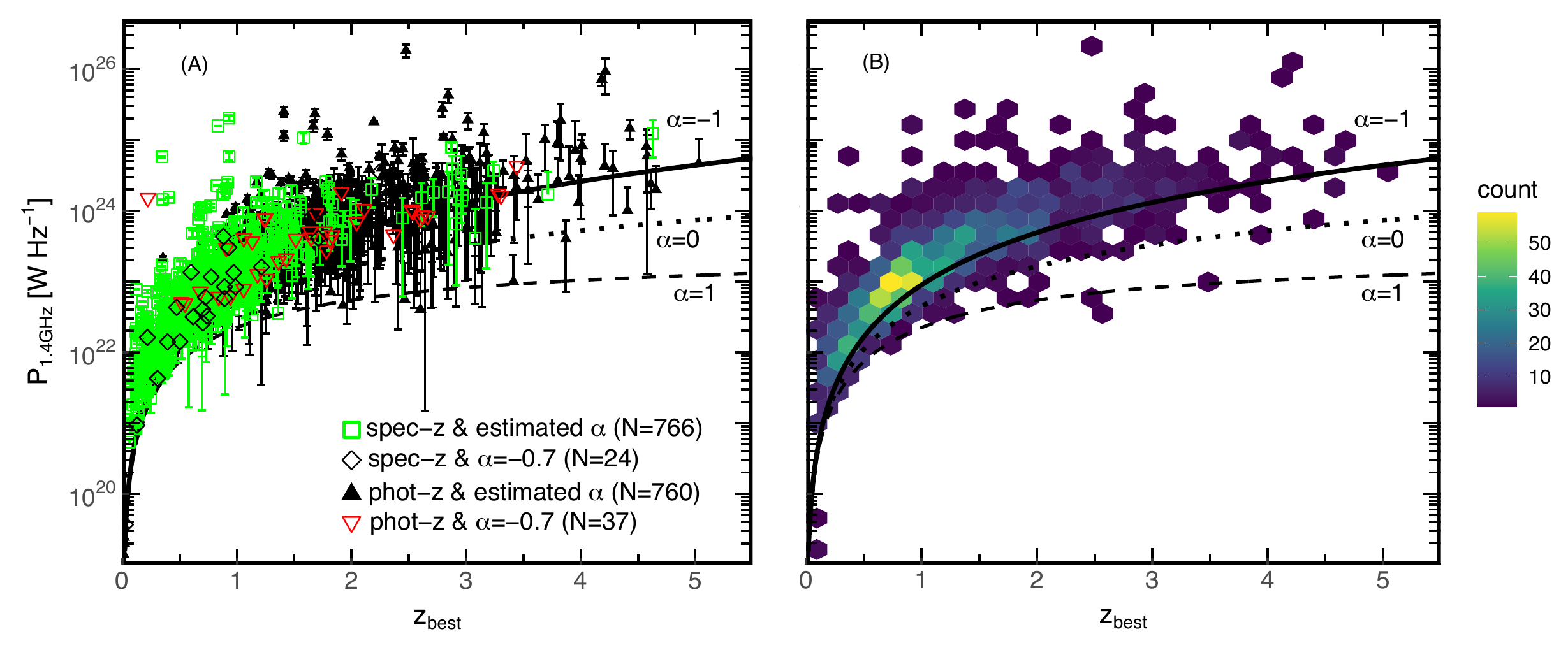}
    \caption{Deroved 1.4~GHz power of individual sources are shown as a function of redshift on the left panel while the same distribution is shown as a hexbin plot on the right panel to reflect the source density better. On the left panel, green open squares and black open diamonds represent spec-z sources with and without estimated spectral index, respectively. Black-filled triangles and red downside triangles represent phot-z sources with and without the estimated spectral index. The three lines delineate $7\sigma$ detection limits for three different assumed spectral index: $\alpha=$-1 (solid), 0 (dotted), and 1 (dashed).}
    \label{fig:power_distr}
\end{figure}

The rest-frame 1.4~GHz radio powers of the sample galaxies computed using their derived spectral indices and redshifts are shown in Figure~\ref{fig:power_distr} along with the survey limits for different spectral indices. Our sample includes 1526 sources (766 spec-z and 760 phot-z) with a measured spectral index while a spectral index $\alpha=-0.7$ is adopted for 61 sources (24 spec-z and 37 phot-z). The $7\sigma$ survey limits for three different assumed spectral indices are also shown: $\alpha=$ -1 (solid line), $\alpha=$ 0 (dotted line), and $\alpha=$ +1 (dashed line). The uncertainty of 1.4~GHz power is mostly attributed to the uncertainty of the spectral index. 

The mean and median 1.4~GHz radio power are $9.87\times10^{23}$ W Hz$^{-1}$ and $1.64\times10^{23}$ W Hz$^{-1}$, respectively. These are characteristic radio powers of star-forming galaxies at $z=1 - 2$.  The right panel of Figure~\ref{fig:power_distr} shows that the large majority of the CHILES Con Pol sources are concentrated approximately along the survey detection limit for $\alpha=-1$, mostly within the radio power range between $10^{22}$ W Hz$^{-1}$ and $10^{24}$ W Hz$^{-1}$. There are 22 sources with $P_{\rm 1.4GHz} \geq 10^{25}$ W Hz$^{-1}$, which is the characteristic radio power of a radio-loud AGN \citep{Miller1990}.  At any given redshift, radio luminous AGNs, distinct from the tight clustering of star forming and Seyfert-like galaxies, are rare.

\subsection{Comparison with Previous Surveys \label{sec:compare}}

\subsubsection{Comparison with MIGHTEE}
MeerKAT observations were conducted on the COSMOS field at 1.4~GHz as part of the MIGHTEE survey \citep{Jarvis2016, Heywood2022, Whittam2024}, which serves as a valuable point of reference for assessing the quality of our source extraction methods. The MIGHTEE survey has published two distinct catalogs, each employing different robust weighting schemes: one with a robust value of $-1.2$ (``$R=-1.2$" hereafter) and another with a value of 0.0 (``$R=0$" hereafter). The $R=-1.2$ image is characterized by a 5\arcsec\ synthesized beam with an RMS noise of 6.0 $\mu$Jy beam$^{-1}$, while the robust $R=0$ image has a synthesized beam size of 8\farcs6 and a thermal noise of $\sigma\sim 1.7$~$\mu$Jy beam$^{-1}$.  Their $R=0$ image is likely suffering from source confusion. \citet{Heywood2022} provided the Level 1 catalog for the $R=0$ image but the Level 0 (raw) catalog for the $R=-1.2$ image\footnote{The flux density in the Level 0 catalog by \citet{Heywood2022} is for 1.22 GHz. Therefore, we converted the flux density by applying a factor of $(1.4/1.22)^{-0.7}$, assuming the mean effective frequency 1.22~GHz and a typical spectral index $-0.7$.}.

\begin{figure}[!ht]
    \centering
    \includegraphics[width=7.1in]{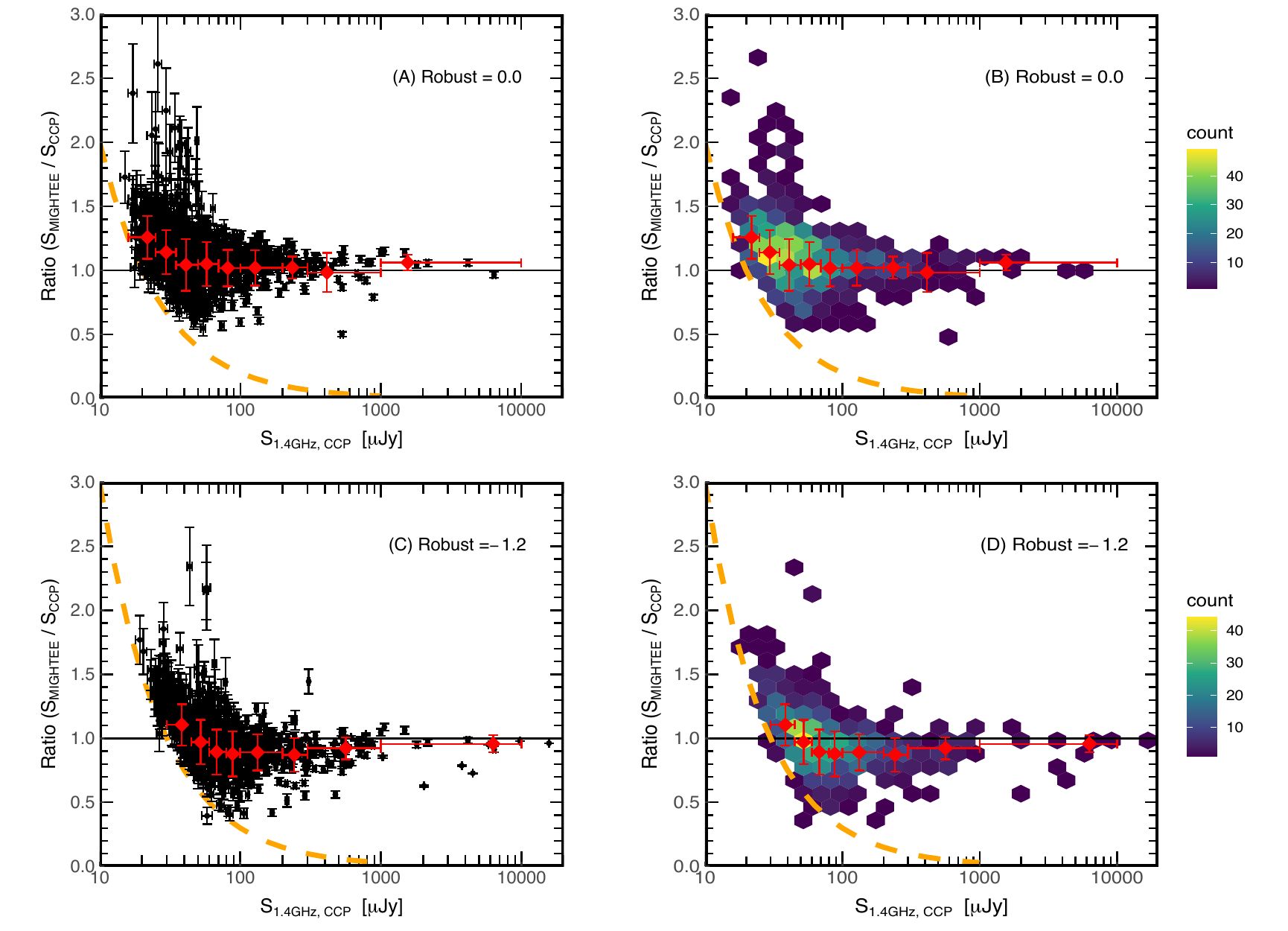}
    \caption{The ratio of flux densities of between MIGHTEE and CHILES Con Pol catalogs with as a function of the CHILES Con Pol 1.4~GHz flux density. Panels (A) and (B) show the comparison of the CHILES Con Pol sources with the $R=0$ MIGHTEE catalog, while panels (C) and (D) present the comparison with the $R=-1.2$ MIGHTEE catalog. Panels (A) and (C) illustrate the ratio of flux densities with their uncertainties, and panel (B) and (D) show the same data as hexbin plots. In panels (A) and (C), red symbols indicate the weighted mean in each bin with uncertainties, and horizontal lines show the range of bins. The orange dashed curves for the $R=0$ catalog (top rows) represent our estimated catalog completeness of 20 \ujy\ ($\sim12\sigma$) while it is 30 \ujy\ ($5\sigma$) for the $R=-1.2$  catalog (bottom rows). }
    \label{fig:comp_mightee}
\end{figure}

The $R=0$ MIGHTEE catalog with a beam size of 8\farcs6 has 1316 sources matching in position with the CHILES Con Pol catalog using a 3\arcsec\ search radius. However, we found that the positional agreement between the two catalogs is not always good, likely because the lower-resolution $R=0$ MIGHTEE sources are not deblended effectively. To reduce uncertainty, we excluded the obvious blended sources in this cross-matching, and a total of 1012 MIGHTEE sources were identified as counterparts to the CHILES Con Pol catalog sources. The comparisons of flux densities between the two catalogs are shown in Figure~\ref{fig:comp_mightee} as the ratio of flux densities in the two catalogs. The ratios, accompanied by their associated uncertainties, are shown in panel (A), while the corresponding hexbin plot is depicted in panel (B) in Figure~\ref{fig:comp_mightee}. There are no matched sources below $S_{\rm 1.4 GHz} = 16$~\ujy\ in the CHILES Con Pol catalog. The weighted mean of the flux density ratios (red diamonds) ranges between $0.98\pm0.09$ and $1.26\pm0.17$ ($0.98\pm 0.09$ to $1.14\pm0.17$ if the lowest flux bin is removed). The measured MIGHTEE flux densities are systematically higher in the lowest flux density bins, reflecting source confusion and noise bias due to a lower angular resolution and incompleteness in the MIGHTEE catalog\footnote{The $R=0$ catalog completeness estimated based on the histogram of source counts as a function of flux density indicates a catalog completeness around 20 \ujy\ ($\sim12\sigma$), which is much higher than expected from the stated RMS noise of the data.  This is also supported by the plot of the flux ratios shown in panels (A) and (B) in Figure~\ref{fig:comp_mightee}.}.  Individual sources with the flux ratio larger than 1.5 are likely blended sources still remaining after removing obvious mismatches (see above).

We also compare the CHILES Con Pol catalog with the $R=-1.2$ MIGHTEE catalog, which has a higher resolution (5\arcsec) and 3.5 times higher noise. Only 725 (43\%) counterparts for the CHILES Con Pol sources are found in the $R=-1.2$ MIGHTEE catalog. The comparisons of flux densities between the CHILES Con Pol and the $R=-1.2$ MIGHTEE catalog are shown in the panels (C) and (D) of Figure~\ref{fig:comp_mightee}. This higher angular resolution MIGHTEE data has many fewer blended sources compared with the $R=0$ catalog, as expected. The weighted mean ratios are between $0.87\pm0.13$ and $1.10\pm0.16$ beyond the flux density of 27.5 \ujy\ ($5\sigma$ for the $R=-1.2$ MIGHTEE catalog). The weighted mean of ratios for $45 \le S_{\rm 1.4GHz, CCP} \le$ 300 \ujy\ is less than 1, indicating that the CHILES Con Pol catalog has systematically larger flux densities than the $R=-1.2$ MIGHTEE catalog. A similar result was found in the comparison of the MIGHTEE catalog and VLA-COSMOS catalog \citep{Heywood2022}, and angular resolution and source confusion appear to have a measurable impact on the flux density derivation in general.

\subsubsection{Comparison with VLA-COSMOS}
\begin{figure}[!h]
    \centering
    \includegraphics[width=7.1in]{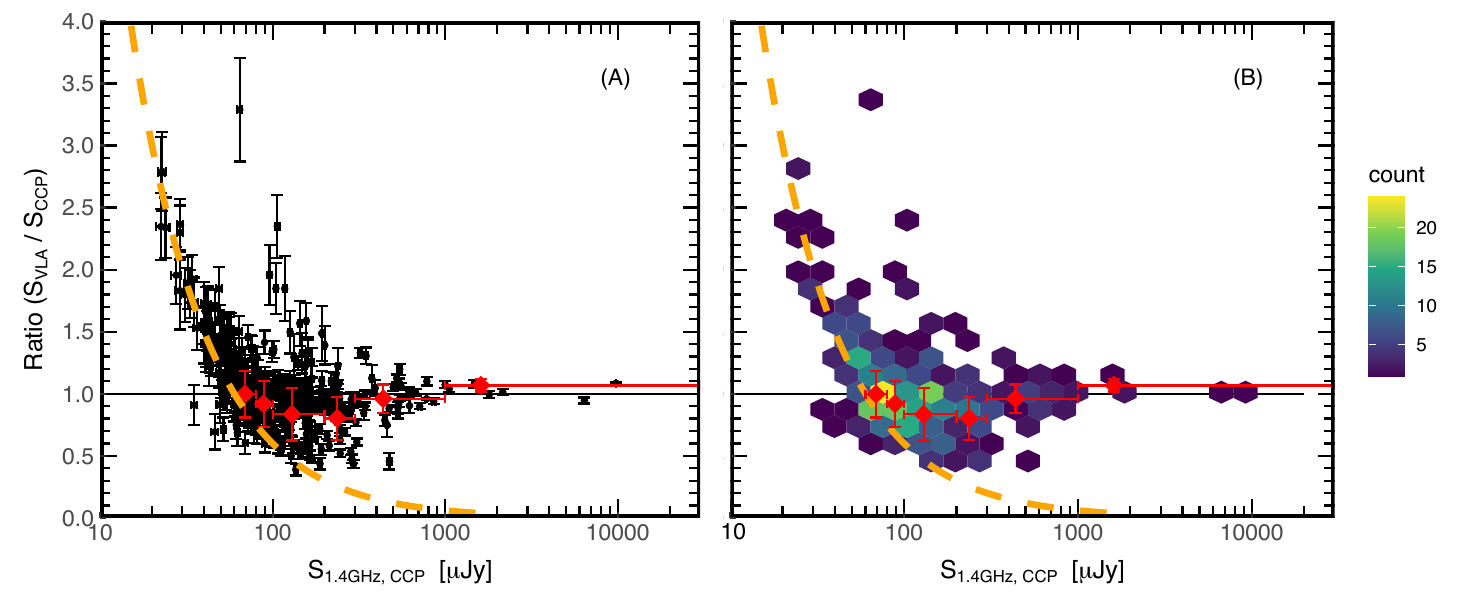}
    \caption{The ratio of flux densities of the VLA-COSMOS sources with respect to those of the CHILES Con Pol sources as a function of CHILES Con Pol 1.4~GHz flux density. The left panel shows the comparison of the CHILES Con Pol with the VLA-COSMOS \citep{Schinnerer2010} and the right panel shows the same data in hexbin plot to include the source density information better. Red diamonds in the left panel indicate the weighted mean in each bin, same as in Figure~\ref{fig:comp_mightee}. The orange dashed curves represent the boundaries of the flux density ratios, determined by the sensitivity limit (60 \ujy\ at $5\sigma$) for the VLA-COSMOS survey. 
    }
    \label{fig:comp_vla}
\end{figure}

The VLA-COSMOS performed the 1.4~GHz radio continuum observations with the VLA in A- and C-configurations, which yielded the synthesized beam size of 2\farcs5 and the RMS noise of $\approx 12$~\ujy\ \citep{Schinnerer2010}. We found 400 counterparts of our CHILES Con Pol in the VLA-COSMOS catalog using a 3\arcsec\ search radius.
The ratio of flux density of VLA-COSMOS to that of CHILES Con Pol is shown in Figure~\ref{fig:comp_vla}. The weighted mean of the ratio is presented as the red diamonds, ranging from $0.80\pm0.17$ to $1.06\pm0.05$. The hexbin plot (right panel) shows that sources are strongly clustered close to the unity ratio. The weighted means of the distributions indicate that the flux densities in CHILES Con Pol catalog is slightly larger than those in the VLA-COSMOS catalog, which might be attributed to the larger beam of the CHILES Con Pol survey (5\farcs5) and possible source blending, similar to the conclusions drawn in the comparisons with the MIGHTEE survey data above.

\subsection{Euclidean-Normalized Differential Source Counts}

We derive the 1.4 GHz radio source counts using the 7$\sigma$ catalog sources within the 50\% sensitivity area of the VLA primary beam. Each source contribution is corrected for completeness (see \S~\ref{sec:completeness}) and an effective survey area for its flux density in the PB-corrected image. The mean flux density ($S_{mean}$) was determined using a weighted mean of total flux density, where the weights were assigned as the inverses of the square of the total flux density uncertainties ($\sigma_{i}$). The Euclidian-normalized differential source density is calculated as:
\begin{equation}
S_{i}^{2.5} {{dN} \over {dS}}_{i} = {{S_{mean,i}^{2.5}} \over {S_{max,i} - S_{min,i}}} \Sigma_{j}^{N} {{1} \over {C_{j} A_{j}}} ~ , 
\end{equation}
where $C_{i}$ is the completeness and $A_{i}$ is the effective area in steradian. 

\begin{figure}[!h]
    \centering
    \includegraphics[width=6.1in]{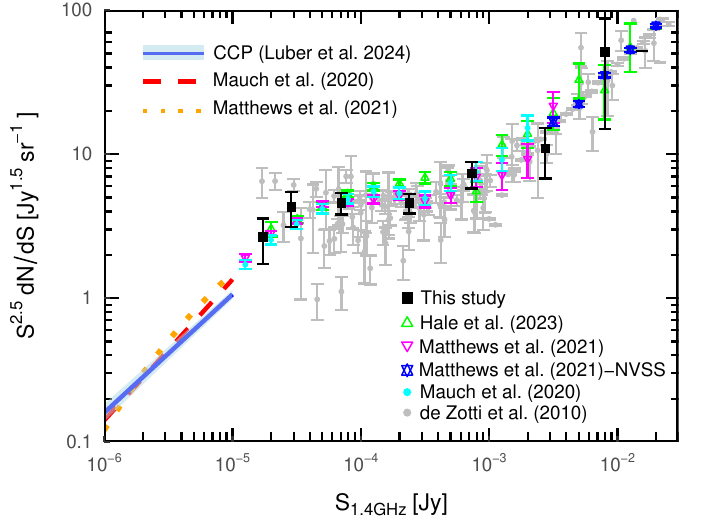}
    \caption{Euclidean-normalized radio source counts including those of the CHILES Con Pol $7\sigma$ catalog sources (black squares).  Published source counts by other recent surveys are also shown for comparison:  \citet{Hale2023} for COSMOS field (green triangles), \citet{Matthews2021b} for DEEP2 field (red downside triangles) and NVSS (blue hexagram), \citet{Mauch2020} for DEEP2 field (blue circles), and \citet{deZotti2010} for the compilation of surveys with gray circles. The models estimated from P(D) analysis are also displayed for comparison: the CHILES Con Pol (Paper 1) in blue solid line, DEEP2 field \citep{Mauch2020} in red dashed line, and NVSS \citep{Matthews2021b} in orange dotted line. }
    \label{fig:dNdS}
\end{figure}

\begin{table}[h]
    \centering
    \caption{Number Count of CHILES Con Pol}
    \begin{tabular}{c|c}
    \hline
        Flux density range & Number Count \\ 
        (\ujy) & (Jy$^{1.5}$ sr$^{-1}$) \\
    \hline
       11.6 -- 20.0 & 2.65$\pm$0.93 \\
       20.0 -- 40.0 & 4.29$\pm$1.17 \\
       40.0 -- 150.0 & 4.58$\pm$0.77 \\
       150.0 -- 500.0 & 4.56$\pm$0.70 \\
       500.0 -- 1500.0 & 7.32$\pm$1.55 \\
       1500.0 -- 6000.0 & 10.97$\pm$4.20 \\
       6000.0 -- 10000.0 & 51.20$\pm$36.30 \\
    \hline
    \end{tabular}
    \label{tab:dNdS}
\end{table}

The Euclidean-normalized differential source counts derived from the CHILES Con Pol survey (black squares) are shown in Figure~\ref{fig:dNdS} along with previous measurements from literature. The source counts derived from the CHILES Con Pol data are generally well aligned  with the findings from the more recent studies, such as MeerKAT observations of the DEEP2 field \citep{Mauch2020, Matthews2021b}, the COSMOS field \citep{Hale2023}, and the 1.4~GHz NRAO VLA Sky Survey observations \citep{Matthews2021b}.  A compilation of earlier 1.4 GHz source counts from literature by \citet{deZotti2010} are also shown. 

Although modern large area surveys benefit from enhanced sensitivity and improved statistics, leading to broad consensus in the source count analyses, systematic variations continue to manifest as subtle but important disparities among the most recent surveys, necessitating careful analysis of multiple contributing factors. For example, our catalog completeness correction is based on simulations of point source recovery. Figure~\ref{fig:resolved} shows that resolved (extended) sources begin to dominate at $S_{\rm 1.4GHz} \ge 42$~\ujy\ for the CHILES Con Pol survey, and our number counts on the brighter regime might be slightly off. The same figure also shows that the CHILES Con Pol source statistics are poor at $S_{\rm 1.4GHz} \ge 500$~\ujy.  As discussed in \S~\ref{sec:compare}, the catalog completeness and source blending can be severe for all surveys, leading to systematic errors at the faintest flux bins.  Additionally, higher angular resolution surveys such as the VLA-COSMOS 3~GHz Large Project \citep{Smolcic2017a} or the VLA 10 GHz survey of the GOODS North field \citep{JimenezAndrade2024} can suffer from poor surface brightness sensitivity and reduced sensitivity to resolved structures, leading to systematic underestimates of source flux density and difficulties in estimating the sample completeness. This issue might happen in our CHILES Con Pol against the MeerKAT observations, as the MeerKAT observations have a slightly better surface brightness sensitivity.

In Figure~\ref{fig:dNdS}, we present the model line obtained from the ``fluctuation analysis'' (or ``P(D)'' analysis) with our CHILES Con Pol image. P(D) analysis was conducted using the CHILES Con Pol image dataset (detailed methodology presented in Paper I). The P(D) distribution was modeled through the convolution of a L\'{e}vy $\alpha$-stable distribution with Gaussian image noise, reflecting the physical reality that the observed P(D) emerges from the convolution of source confusion and instrumental noise components. The L\'{e}vy $\alpha$-stable distribution provides an analytically optimal framework for characterizing the pixel distribution arising from confusion noise generated by unresolved sources (Herranz et al. 2004). This distribution is fully parameterized by four variables: $\alpha$ (the characteristic exponent governing the function's impulsiveness), $\beta$ (the skewness parameter), $\gamma$ (the scale parameter controlling the distribution's spread), and $\mu$ (the distribution's location parameter). Under the assumption of maximum skewness ($\beta=1$), a Markov Chain Monte Carlo (MCMC) optimization procedure was implemented to determine the optimal parameters, including three parameters of the L\'{e}vy $\alpha$-stable distribution and the RMS of the Gaussian noise ($\sigma_{0}$). The analysis yielded a best-fit differential number count relationship expressed as
\begin{equation} 
S^{2.5} \frac{dN}{dS} = 13746 S^{0.823} 
\end{equation}
, where the best-fit parameters were obtained by analyzing the pixel distribution in the flux density range of -10 $< S <$ 10~\ujy\ (see Paper I for more detail). 
Our model line (in blue solid line) is generally consistent with the P(D) analyses by \citet{Mauch2020} and \citet{Matthews2021b}, and extend the down-turn in the number counts below 50 \ujy\ traced by individually detected sources. The P(D) analysis by the CHILES Con Pol survey is slightly flatter than the other surveys. As discussed in Paper I, both the slope and normalization of the CHLES Con Pol results depend on the range of pixel values included in the analysis, possibly indicating a curvature in the actual number counts above and below flux density of 10 \ujy.

\section{Conclusion} \label{sec:summary}

A continuum source catalog of the CHILES Con Pol (Paper 1) derived using two different source extraction programs, Blobcat \citep{Hales2012} and PyBDSF \citep{Mohan2015}, is presented and discussed. A total of 1678 radio continuum sources above 7$\sigma$ ($\sigma=1.67$ \ujy\ beam$^{-1}$) have been identified from the Stokes I image with the effective frequency of 1.447~GHz. For the synthesized beam of 5\farcs5$\times$5\farcs0 (FWHM), 772 sources are classified as resolved, while 906 are considered unresolved. A completeness study demonstrated that our catalog is 90\% complete at $9\sigma$ and 96\% complete at 15$\sigma$. 

A total of 790 spectroscopic redshifts have been compiled from the literature, and a photometric redshift has been assigned for the remaining 797 CHILES Con Pol sources from COSMOS 2020-R1 catalog \citep{Weaver2022}.  These redshifts are used to compute 1.4~GHz power $P_{1.4GHz}$, utilizing the derived radio spectral index. Over one-half of all radio sources are located in $z<1$, and a possible overdensity of radio sources is identified at $z \sim 2.6$. 

Radio spectral index of the 1614 (96.2\%) CHILES Con Pol sources are computed from the four SPW images (covering 1.063 -- 1.831~GHz) assuming a power-law model, with all beam sizes matched to the largest beam in the lowest frequency image. We found empirically that deriving a reliable spectral index measurement requires a total S/N of $>20$. Our derived spectral index distribution peaks at $\alpha=-0.706$, with a small secondary peak near $\alpha\sim0$. The spectral index distribution shows a significant spread spanning between $\alpha=1$ and $\alpha=-1.5$, even at S/N $\ge 20$.  We examine the relationships between radio spectral index and radio power, as well as between radio spectral index and redshift, to explore the intrinsic spread of the spectral index distribution. but no significant correlations are found. 

A comparison of the CHILES Con Pol survey with the MIGHTEE \citep{Heywood2022} and VLA-COSMOS \citep{Schinnerer2010} surveys show that our flux density measurements are consistent with those of the previous surveys at high flux densities, but a significant completeness problem for the published catalogs from these comparison surveys is also clearly seen. Our analysis of the Euclidean-normalized source counts show a good agreement among these recent surveys.  Small discrepancy among these surveys as well as sometimes severe discrepancies of earlier surveys can be understood in terms of systematic problems with survey and catalog completeness.   A detailed discussion on the nature of the faint radio sources detected in the CHILES Con Pol survey will be presented in the forthcoming paper (Gim et al. in prep.).

\facility{VLA}

\software{Blobcat \citep{Hales2012}, PyBDSF \citep{Mohan2015}, R \citep{rcite}, CARTA \citep{carta}}

\section*{Acknowledgement}
The authors gratefully acknowledge valuable discussions and input from Dr. Jacqueline van Gorkom. We also thank Dr. Urvashi Rau at NRAO for her insightful comments, and Dr. Christopher Hales at NRAO for his contributions to the initiation of this project. In addition, the authors are grateful to the referee for a valuable inputs. MSY's research is partially supported by the NSF grant AST 1413102. DJP greatly acknowledges support from the South African Research Chairs Initiative of the Department of Science and Technology and National Research Foundation.

\bibliographystyle{aasjournal}

\end{document}